\documentclass[aps,prb,twocolumn,superscriptaddress,showpacs]{revtex4-1}

\usepackage{graphicx}

\begin{document}

\newcommand{\Fref}[1]{Figure~\ref{#1}}
\newcommand{\fref}[1]{Fig.~\ref{#1}}
\newcommand{\etal}{\emph{et al.}}
\newcommand{\tc}{$T_c \cong$}
\newcommand{\srpdge}{SrPd$_2$Ge$_2$}

\title{Magnetic penetration depth in single crystals of SrPd$_2$Ge$_2$ superconductor}

\author{H. Kim}
\email{hyunsoo@iastate.edu}
\affiliation{The Ames Laboratory and Department of Physics and Astronomy, Iowa State University, Ames, IA 50011, USA}

\author{N. H. Sung}
\email{nakheon@gmail.com}
\affiliation{School of Materials Science and Engineering, Gwangju Institute of Science and Technology (GIST), Gwangju 500-712, Korea}

\author{B. K. Cho}
\email{chobk@gist.ac.kr}
\affiliation{School of Materials Science and Engineering, Gwangju Institute of Science and Technology (GIST), Gwangju 500-712, Korea}
\affiliation{Department of Photonics and Applied Physics, Gwangju Institute of Science and Technology (GIST), Gwangju 500-712, Korea}

\author{M. A. Tanatar}
\email{tanatar@ameslab.gov}
\affiliation{The Ames Laboratory and Department of Physics and Astronomy, Iowa State University, Ames, IA 50011, USA}

\author{R. Prozorov}
\email[Corresponding author: ]{prozorov@ameslab.gov}
\affiliation{The Ames Laboratory and Department of Physics and Astronomy, Iowa State University, Ames, IA 50011, USA}

\date{25 January 2013}

\begin{abstract}
The in-plane magnetic penetration depth, $\lambda_m(T)$, was measured in single crystals of SrPd$_2$Ge$_2$ superconductor in a dilution refrigerator down to $T=60$ mK and in magnetic fields up to $H_{dc} = 1$ T by using a tunnel diode resonator. The London penetration depth, $\lambda$, saturates exponentially approaching $T\rightarrow 0$ indicating fully gapped superconductivity. The thermodynamic Rutgers formula was used to estimate $\lambda(0) = 426$ nm which was used to calculate the superfluid density, $\rho_s(T)=\lambda^2(0)/\lambda^2(T)$. Analysis of $\rho_s(T)$ in the full temperature range shows that it is best described by a single - gap behavior, perhaps with somewhat stronger coupling. In a magnetic field, the measured penetration depth is given by the Campbell penetration depth which was used to calculate the theoretical critical current density $j_c$. For $H \le 0.45$ T, the strongest pinning is achieved not at the lowest, but at some intermediate temperature, probably due to matching effect between temperature - dependent coherence length and relevant pinning lengthscale. Finally, we find a compelling evidence for surface superconductivity. Combining all measurements, the entire $H$-$T$ phase diagram of SrPd$_2$Ge$_2$ is constructed with an estimated $H_{c2}(0)=0.4817$ T.
\end{abstract}


\maketitle

\section{Introduction}

Superconductivity in the tetragonal ThCr$_2$Si$_2$-type SrPd$_2$Ge$_2$ was discovered first in polycrystalline \cite{Fujii2009} and later in single crystals \cite{Sung2011} with the superconducting phase transition temperature ($T_c$) at 3.0 K and 2.7 K, respectively. The upper critical field ($H_{c2}$) was estimated to be 4920 Oe at $T=0$ by using Helfand-Werthamer (HW) theory \cite{Helfand1966} based on the experimental data obtained only down to $T=0.7T_c$.\cite{Sung2011} It has been found that $T_c$ and $H_{c2}$ can be slightly increased by chemical doping.\cite{Sung2012} The London penetration depth and coherence length are reported to be $\lambda(0)=566$~nm and $\xi(0)=$~21~nm \cite{Kim2012} and $\lambda(0)=345 \pm 30$~nm $\xi(0)=25.6\pm$~0.5~nm.\cite{Samuely2013} These values give the Ginzburg - Landau parameter of $\kappa = 27$ \cite{Kim2012} and $\kappa = 13.5$ \cite{Samuely2013}, which makes SrPd$_2$Ge$_2$ a strong type-II superconductor. Furthermore, thermodynamic \cite{Sung2011} and tunneling spectroscopy measurements  are consistent with a slightly strong - coupling s-wave Bardeen-Cooper-Schrieffer (BCS) superconductor with the zero-temperature value of the superconducting gap of $\Delta_0 \approx 2k_BT_c$\cite{Kim2012,Samuely2013}, - not far from the weak coupling value of 1.76.

This superconductor is interesting particularly because of compositional similarity to the newly discovered isostructral Fe-and Ni-pnictide superconductors with comparable $T_c$ such as KFe$_2$As$_2$, BaNi$_2$As$_2$, and SrNi$_2$P$_2$. Although there is strong experimental evidence for nodal superconductivity in KFe$_2$As$_2$ (Ref. \onlinecite{Hashimoto2010,Reid2012}), the Ni-based ones have been shown to be fully gapped by thermodynamic and thermal transport measurements.\cite{Kurita2009,Kurita2011} This naturally prompts the question: what the structure of a superconducting gap is in SrPd$_2$Ge$_2$? So far, not much work has been done on SrPd$_2$Ge$_2$ in this direction. Tunneling spectroscopy between 0.17$T_c$ and $T_c$ is consistent with a single, isotropic gap superconductor.\cite{Kim2012} To verify this, however, the thermodynamic, thermal transport, and penetration depth measurements down to much lower temperatures are necessary to provide objective conclusions regarding the gap-symmetry in SrPd$_2$Ge$_2$.

The magnetic penetration depth in superconducting state is among most useful probes to explore the superconducting state.\cite{Prozorov2006} In zero external magnetic field, it is defined by the London penetration depth. If measured with sufficient accuracy and low enough temperatures, it can be used to understand the angular variation of the superconducting gap on the Fermi surface.\cite{Prozorov2006,ProzorovKoganROPP2011}
In the presence of vortices, the magnetic penetration depth is also influenced by the Campbell penetration depth which depends on the elastic properties of vortex lattice and is linked directly to the critical current density.\cite{Brandt1995} There has been only limited work performed in the mixed state of SrPd$_2$Ge$_2$ \cite{Sung2012a,Samuely2013} and no studies of the critical current density over the full temperature and field range.

Here we report precision tunnel diode resonator measurements of the magnetic penetration depth, $\lambda_m(T)$, in single crystals of SrPd$_2$Ge$_2$ ($T_c=2.7$ K) performed in a dilution refrigerator with temperatures down to $T\approx 0.02T_c$ and in magnetic fields up to 1 T $\approx 2H_{c2}(0)$. The low - temperature variation of the London penetration depth, $\Delta\lambda (T)$, clearly indicates exponential saturation, and the analysis of the full - temperature range superfluid density is consistent with a fully gapped superconductor. The thermodynamic Rutgers formula was used to estimate $\lambda(0) = 426$ nm which was used to calculate the superfluid density, $\rho_s(T)=\lambda^2(0)/\lambda^2(T)$. Analysis of $\rho_s(T)$ in the full temperature range shows that it is best described by a single - gap behavior, perhaps with somewhat stronger coupling. The upper critical field, $H_{c2}(0) = 0.4817$~T was determined by the HW theory from the field - sweeps at different temperatures down to 0.02$T_c$. In finite magnetic fields, $H < H_{c2}(0)$, the Campbell penetration depth, $\lambda_C(T,H)$, shows minimum at an intermediate temperatures (rather than at the lowest temperature) which indicates non-monotonic variation of the theoretical critical current ($j_c$) calculated from $\lambda_C(T,H)$. Additional diamagnetic response detected above $H_{c2}(T)$ is consistent with surface superconductivity with $H_{c3}(T)=1.695H_{c2}(T)$.\cite{Tinkham2004}

\section{experimental}

Single crystals of SrPd$_2$Ge$_2$ were grown using a self-flux method as described in Ref.~\onlinecite{Sung2011}. The magnetic penetration depth was measured in a dilution refrigerator by using a tunnel diode resonator (TDR) technique (for review, see Ref. \onlinecite{Prozorov2006}). The sample with dimensions (0.51$\times$0.70$\times$0.04) mm$^3$ with the shortest direction being along the $c-$axis was mounted on a sapphire rod and inserted into a 2 mm inner diameter copper coil that produces rf excitation field with empty-resonator frequency of 17 MHz with amplitude $H_{ac} \sim 20$ mOe, much smaller than $H_{c1}$ of typical conventional superconductors. Measurements of the in-plane magnetic penetration depth were done with both $H_{dc}$ and $H_{ac} \parallel c$-axis. The shift of the resonant frequency (in cgs units), $\Delta f(T)=-G4\pi\chi(T)$, where $\chi(T)$ is the differential magnetic susceptibility, $G=f_0V_s/2V_c(1-N)$ is a constant, $N$ is the demagnetization factor, $V_s$ is the sample volume and $V_c$ is the coil volume. The constant $G$ was determined from the full frequency change by physically pulling the sample out of the coil. With the characteristic sample size, $R$, $4\pi\chi=(\lambda/R)\tanh (R/\lambda)-1$, from which $\Delta \lambda$ can be obtained.\cite{Prozorov2000,Prozorov2006}

\section{Results and discussion}

\subsection{London penetration depth}

\begin{figure}[tbh]
\includegraphics[width=1.0\linewidth]{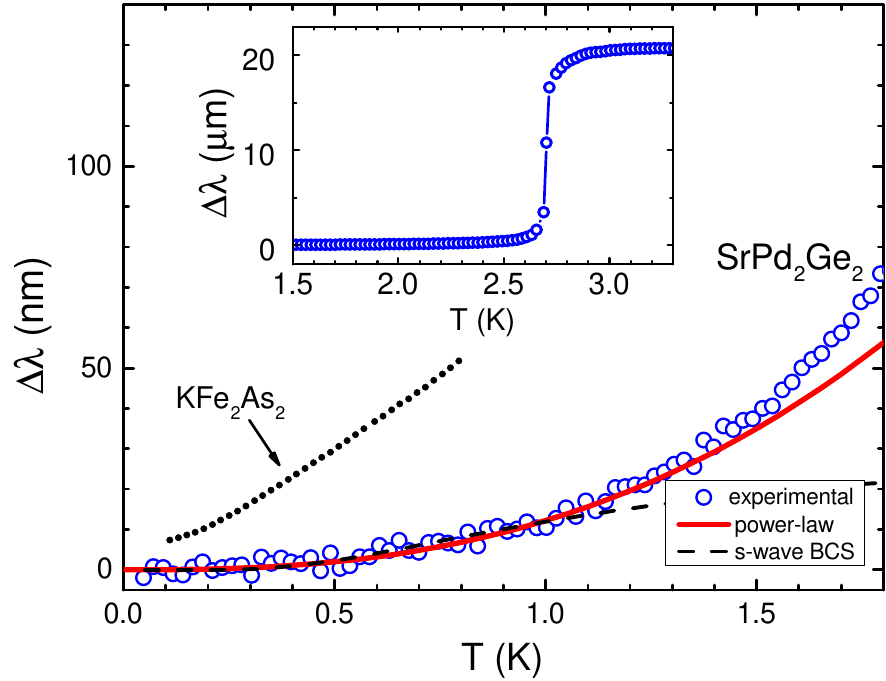}%
\caption{\label{fig:lambda}
In-plane London penetration depth in a single crystal of SrPd$_2$Ge$_2$. Main panel: Open circles represent experimental data. Solid and dashed lines represent power-law and BCS (single gap s-wave) low - temperature fitting. Dotted shows the data for KFe$_2$As$_2$ taken from Ref.~\onlinecite{Hashimoto2010} for comparison. Inset: London penetration depth in the full temperature range demonstrating a sharp transition at $T_c=2.7$ K}.
\end{figure}

Figure~\ref{fig:lambda} shows temperature variation of the in-plane London penetration depth, $\Delta\lambda(T)$, measured in a single crystal of SrPd$_2$Ge$_2$ superconductor which exhibits a very sharp superconducting phase transition at $T_c=2.7$ K as shown in the inset, indicating a high quality, homogeneous sample. In the main panel, $\Delta\lambda(T)$ is shown with temperatures up to about $0.67 T_c$. The saturation in $T\rightarrow 0$ limit and almost flat temperature dependence, $\Delta\lambda(T_c/3)< 10$ nm, indicate fully gapped superconductivity. Experimental $\Delta\lambda(T)$ can best fit to a power-law function, $\Delta\lambda(T)=AT^n$, with the exponent of $n=2.7\pm 0.1$ and pre-factor of $A=12.2\pm 0.4$ nm/K$^{2.7}$. The fitting curve is shown in red solid line. A power-law function with such a high exponent has very weak variation at low temperatures, indistinguishable from the exponential behavior which is predicted for a fully open superconducting gap. In fact, the BCS low - temperature form, $\Delta\lambda(T)=\lambda(0)\sqrt{\pi\Delta_0/2k_BT}\exp{(-\Delta_0/k_BT)}$, where $\Delta_0$ is the maximum gap value at $T=0$, fits the data equally well for $T<T_c/3$ where it is expected to be valid. However, the best fitting is achieved with $\lambda(0)=50$ nm and $\Delta_0=0.74k_BT_c$. The latter is impossible in the single - gap clean limit where $\Delta_0\approx 1.76k_BT_c$ is expected. The value of $\lambda(0)$ is also much smaller than the reported value of 566 nm.\cite{Kim2012} Similar low - temperatures features can be seen in two - band superconductors such as MgB$_2$ (Ref. \onlinecite{FletcherMgB2}), 2H-NbSe$_2$ (Ref. \onlinecite{Fletcher2007NbSe2}), Lu$_2$Fe$_3$Si$_5$ (Ref. \onlinecite{Gordon2008}), and more recently LiFeAs (Ref. \onlinecite{Kim2011}). However, as we show below, analysis of the superfluid density in the full temperature range is inconsistent with a two - gap clean behavior. Instead, it is more likely that we are dealing with moderate pair - breaking scattering (maybe due to well-known magnetic impurities in Pd) which results in a finite density of states inside the gap. We also point out that these temperature variation is very small compared to a known nodal superconductor KFe$_2$As$_2$ (Ref. \onlinecite{Hashimoto2010,Reid2012}) with similar $T_c$, but exhibiting much stronger temperature dependence of $\Delta\lambda$, indicating significant amount of quasiparticles generated at the low temperatures, most likely due to nodes in the gap.

For a metallic sample, the measured penetration depth above $T_c$ is determined either by the skin depth $\delta$ or sample size. In case of skin depth limiting, the value of $\lambda(T>T_c)$ shown in the inset in Fig.~\ref{fig:lambda} is one half of the actual skin depth.\cite{Lifshitz1984} Therefore, we can estimate normal state resistivity from the measurements using $\rho=(2\pi \omega/c^2)\delta^2$.\cite{Kim2011} For SrPd$_2$Ge$_2$ with $\omega/2\pi=f_0= 17$ MHz and $\delta /2\approx 20$ $\mu$m, the calculated resistivity is approximately 12 $\mu \Omega$ cm which is much less than the experimental value of 68 $\mu\Omega$ cm (Ref.~\onlinecite{Sung2011}). Therefore, we conclude that it is in a sample - size limited regime. Using the same equation, the estimated skin depth is 180 nm which is close to half width of the sample.

Finally, we note that the data exhibit a smooth transition from superconducting penetration depth to the normal state between $T=T_c$ and $T^*\approx 3.0$ K, which has also been seen in transport measurement.cite{Sung2011} Interestingly, $T^*=3$ K is the onset of superconductivity observed in a polycrystalline sample.\cite{Fujii2009} Similar feature has also been observed in a related superconductor BaNi$_2$As$_2$.\cite{Ronning2008} Perhaps this feature requires further study.

\subsection{Superfluid density}
\begin{figure}[tbh]
\includegraphics[width=1.0\linewidth]{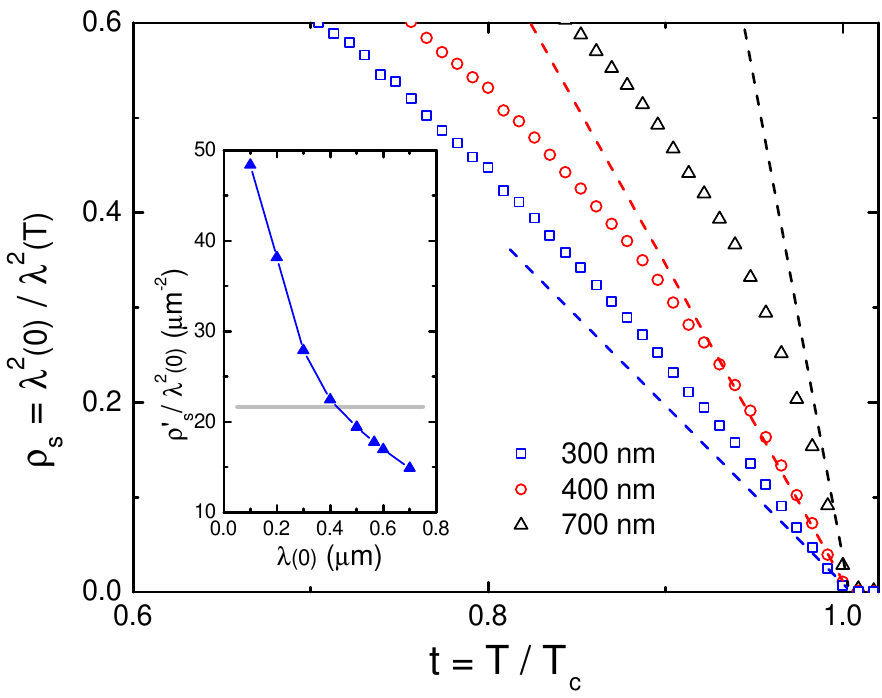}%
\caption{\label{fig:Rutgers} Main panel: Calculated superfluid density, $\rho_s$, with various $\lambda(0)$'s. The slope of dashed lines was determined with $\Delta C_p$ and $|dH_{c2}/dT|_{T_c}$ using the Rutgers formula as described in the text. Inset: Variation of $|\partial \rho_s/\partial t|_{T_c}/\lambda^2(0)$ with varying $\lambda(0)$. Here $\rho_s'$ was determined from the experimental data. The gray band is a theoretical estimate with 5\% hypothetical error in $\Delta C_p$ and $|dH_{c2}/dT|_{T_c}$.}
\end{figure}

While the low - temperature behavior is important, the superconducting gap can be probed at all energy scales by the analysis the superfluid density, $\rho_s(T)=\lambda^2(0)/\lambda^2(T)$, in the entire temperature range.\cite{ProzorovKoganROPP2011} However, superfluid density requires knowledge of the absolute value of $\lambda(0)$. For \srpdge~ $\lambda(0)=566$ nm was estimated using a dirty limit \cite{Kim2012} and $\lambda(0)\approx 390$ nm was extracted from field - dependent magnetization \cite{Sung2011} within the London approximation and $\lambda(0)\approx 345 \pm5$ nm was estimated from the measurements of the field of first penetration.\cite{Samuely2013} So, the variation of the literature values is quite significant and we have to resort to another,  thermodynamic, approach based on the Rutgers formula. In the Ginzburg-Landau regime, i.e. near $T_c$, it can be shown that

\begin{equation}\label{eq:Rutgers}
\left|\frac{\partial \rho_s}{\partial t}\right|_{T_c}=\frac{16\pi^2\lambda^2(0)}{\phi_0| \partial H_{c2}/\partial T|_{T_c}} \Delta C_p
\end{equation}

\noindent where $\phi_0=2.07\times 10^{-7}$ G cm$^2$ is a flux quantum and $|\partial H_{c2}/\partial T|_{T_c} =0.26$ T/K is determined experimentally (see Fig.~\ref{fig:HT}). Specific heat jump $\Delta C_p=7381$~erg/cm$^3$K is taken from Ref.~\onlinecite{Sung2012}. Applying these thermodynamic values suggests $|\partial \rho/\partial t|_{T_c}/\lambda^2(0)=21.7$ $\mu$m$^{-2}$ where $t=T/T_c$ is the reduced temperature. This quantity can be compared with the actual slope of calculated $\rho_s(t)$ with various $\lambda(0)$ at $T_c$ as shown in Fig.~\ref{fig:Rutgers}. In the main panel, the open symbols represent the calculated superfluid density with 300, 400, and 700 nm in triangle, circle, and square, respectively. The dashed lines are determined with the slope calculated by Eq.~\ref{eq:Rutgers} for three values of $\lambda(0)$ quoted above. The line with 400 nm shows very good agreement with calculated $\rho_s$ while the line with 300 nm significantly underestimates, and the one with 700 nm overestimates. This procedure can be repeated with various values of $\lambda(0)$. The result is summarized in the inset where the solid triangle represents experimental slopes obtained by fitting experimental data near $T_c$ to a linear line. The gray horizontal band represents the theoretical value of $|\partial \rho/\partial t|_{T_c}/\lambda^2(0)=21.7\pm 2.2$ $\mu$m$^{-2}$ determined with a 5\% hypothetical error in $|\partial H_{c2}/\partial T|_{T_c}$ and $\Delta C_p$. In this way, $\lambda(0)$ can be determined at the intersection of the theoretical line and experimental results, which provides that $\lambda(0)=426\pm 60$ nm that lies between the literature values. With this value, the slope of $\rho_s$ at $T_c$ is determined to be $-3.9$.

\begin{figure}[tbh]
\includegraphics[width=1.0\linewidth]{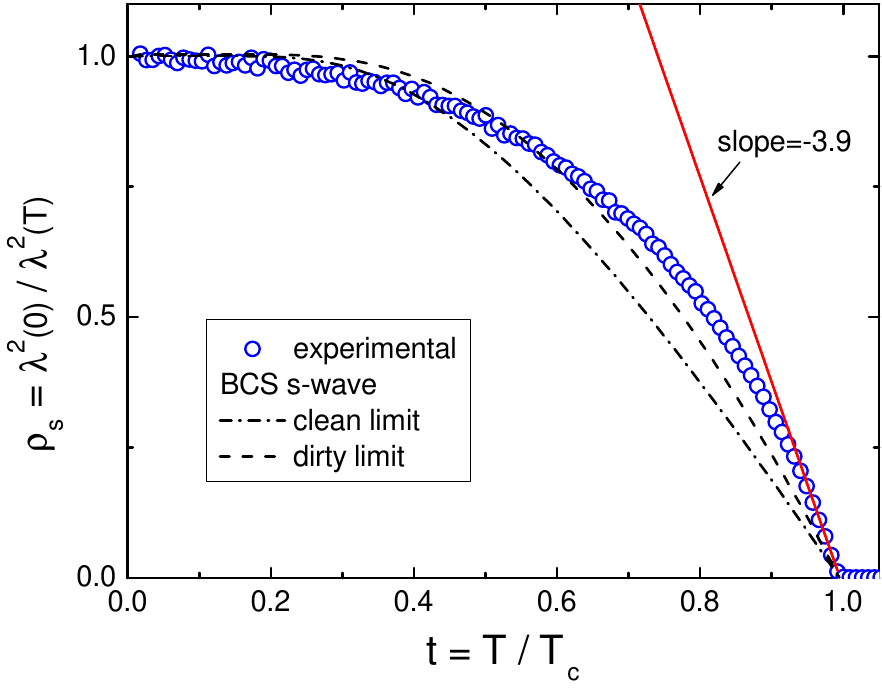}%
\caption{\label{fig:rho} Calculated superfluid density, $\rho_s(T)=\lambda^2(0)/\lambda^2(T)$ using $\lambda(0)=426$ nm. Open circles represent the experimental data. The dashed dots and dashed lines represent single - gap weak - coupling s-wave BCS superconductor in clean and dirty limit, respectively.}
\end{figure}

The calculated superfluid density with $\lambda(0)=426$ nm is shown in Fig.~\ref{fig:rho}. The dot - dashed and dashed lines show expectation for clean and dirty limit of a single-gap BCS superconductor in the weak - coupling limit, respectively. An attempt to use a two - gap (clean) $\gamma-$ model \cite{Kogan2009} in the full - temperature range converges to a single - gap limit with $\Delta(0)/k_BT_c =$2.2. Therefore, the gap symmetry of \srpdge~ is most likely represented by a single gap s-wave, perhaps with somewhat enhanced coupling strength. It was noted previously that the shape of $\rho_s(T)$ is rather close to a nonlocal - limiting case, expected in type I superconductors such as aluminum and cadmium.\cite{Bonalde2003} Similar argument was made in the work by T.~K.~Kim {\it et al.} in which SrPd$_2$Ge$_2$ appeared to be type I according to the intrinsic electronic structure despite the fact that experimental $\xi(0)$ and $\lambda(0)$ put it in a strong type-II regime.\cite{Kim2012} In any case, our study confirms that simple analysis with an isotropic Fermi surface is not sufficient and, perhaps, the results could be explained by taking into account a realistic band structure. We can, however, conclude that no nodes are present in the superconducting gap.

\subsection{Campbell penetration depth}

\begin{figure}[tbh]
\includegraphics[width=1.0\linewidth]{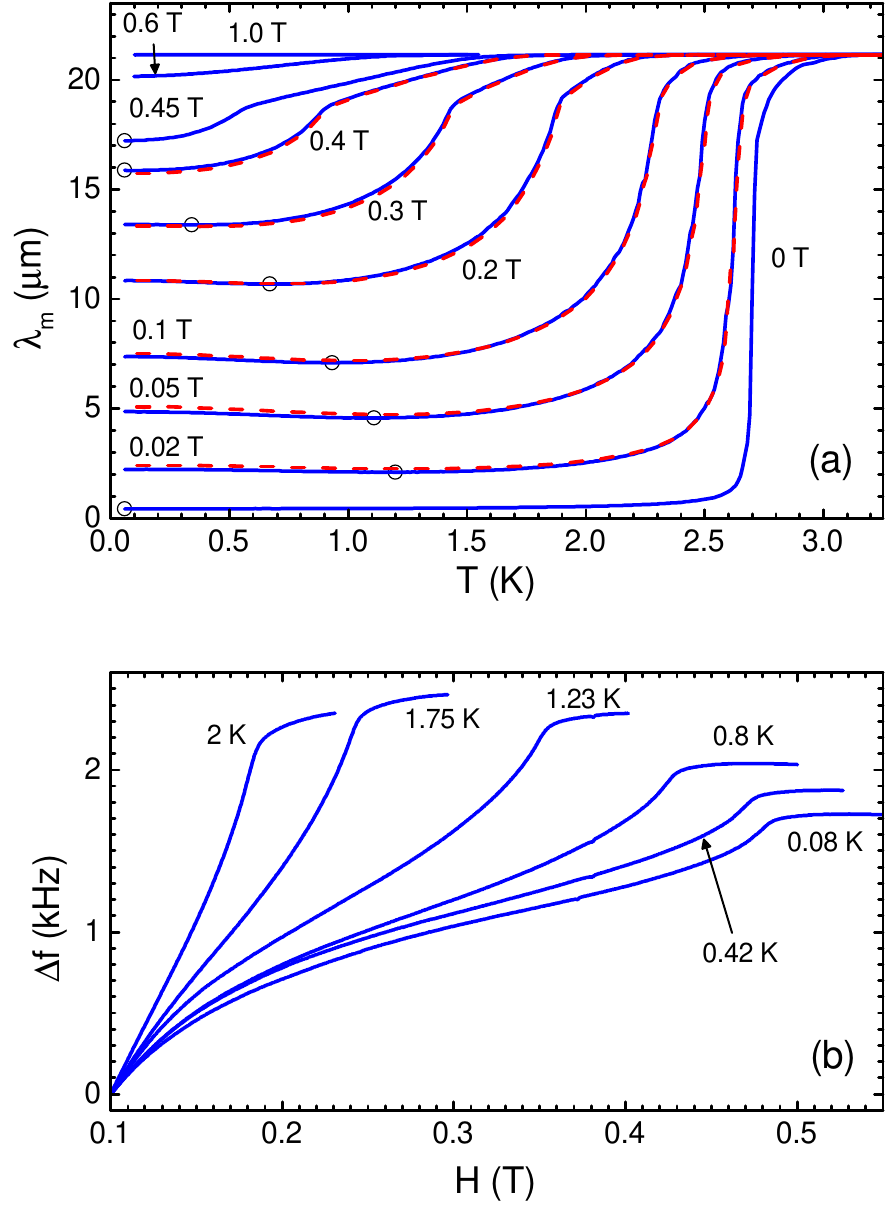}%
\caption{\label{fig:infield}(a) Zero-field cooled (ZFC) and field cooled (FC) in-plane magnetic penetration depth measured in a single crystal of SrPd$_2$Ge$_2$ shown by solid and dashed lines, respectively. Open circles mark the minimum in $\lambda_m(T)$ in each magnetic field. (b) Magnetic field sweeps at fixed temperatures.}
\end{figure}
%

\begin{figure}[tbh]
\includegraphics[width=1.0\linewidth]{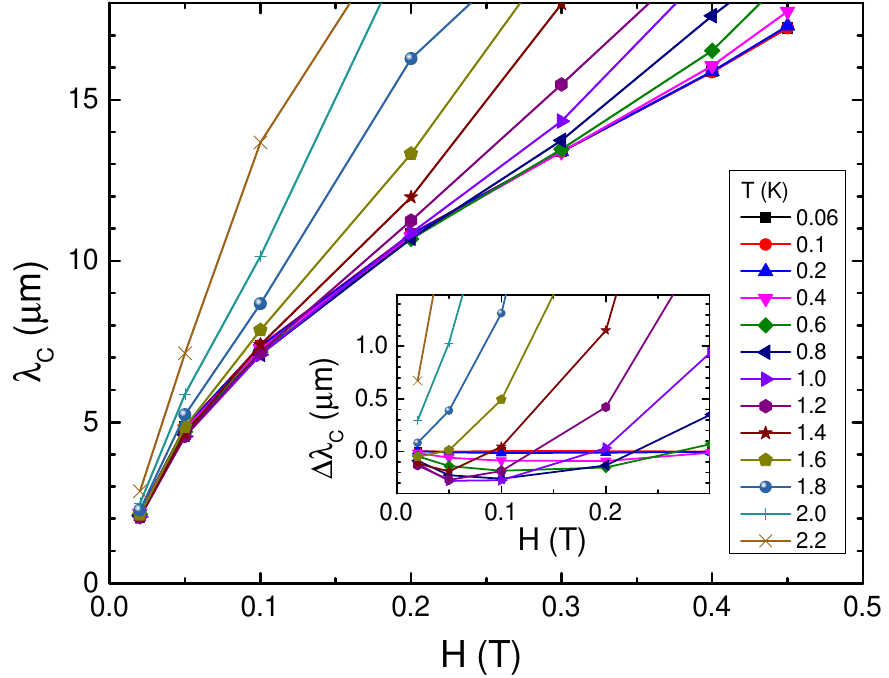}%
\caption{\label{fig:Campbell}Isothermal Campbell penetration depth as function of a DC magnetic field calculated from the data of Fig.~\ref{fig:infield}(a). Inset: $\Delta\lambda_C(T,H)=\lambda_C(0.06 \textmd{ K},H)-\lambda_C(T,H)$ is displayed to show non-monotonic behavior.}
\end{figure}

Figure~\ref{fig:infield}(a) shows magnetic penetration depth, $\Delta\lambda_m(T,H)$, as a function of temperature and magnetic field measured after cooling without magnetic field to target low temperature and then applying a DC magnetic field of indicated amplitude (ZFC - solid lines) and upon cooling in field (FC - dashed lines). Increasing DC magnetic field not only suppresses the superconducting phase transition and diamagnetic shielding, but also induces another diamagnetic phase between the normal state and apparent bulk superconductivity. This feature appears much clearer above $H=0.2$~T and persists at least up to $H=0.6$~T which is far greater than the upper critical field of the bulk superconductivity. This can hardly be understood with sample inhomogeneity or second phases since they both should affect the measurements in zero field. Systematic tracking of this feature reveals possible connection to surface superconductivity, which will be discussed later together with the general $H$-$T$ phase diagram. In addition to temperature sweeps at fixed DC magnetic fields, magnetic field sweeps at different fixed temperatures were performed, and superconducting transitions are clearly detected as shown in Figure~\ref{fig:infield}(b). The determined bulk $H_{c2}(T)$ is consistent with $T_c(H)$ determined from the temperature scans. We note that the anomaly above $H_{c2}(T)$ was not detected in the field sweeping measurements.

Another unusual experimental observation is apparently non - monotonic temperature variation of $\Delta\lambda_m(T,H)$ for 0.02 T $\le H \le $ 0.4 T in both ZFC and FC data. The minima of $\Delta\lambda_m(T,H)$ are marked by open circles for clarity in Fig.~\ref{fig:infield}(a), and the locations of these minima are nearly the same for both ZFC and FC data. This feature is too shallow to be due to paramagnetic impurities (and should be most pronounced at $H=$0).\cite{Cooper1996} We suggest that this behavior is cause by the size - matching of the temperature - dependent coherence length and relevant pinning centers. Furthermore, we observed the inversion of diamagnetic screening between ZFC and FC runs at about $H=0.2$ T. The only way to obtain a difference between FC and ZFC experiments in AC response is to assume a non - parabolic pinning potential \cite{Prozorov2003} and our results would indicate that the effective pinning potential shape is strongly field and temperature dependent. Conventional superconductors, Al (Ref. \onlinecite{KimAl}) or YbSb$_2$ (Ref. \onlinecite{Zhao2012}) do not exhibit such upturns in a magnetic field.

\begin{figure}[tbh]
\includegraphics[width=1.0\linewidth]{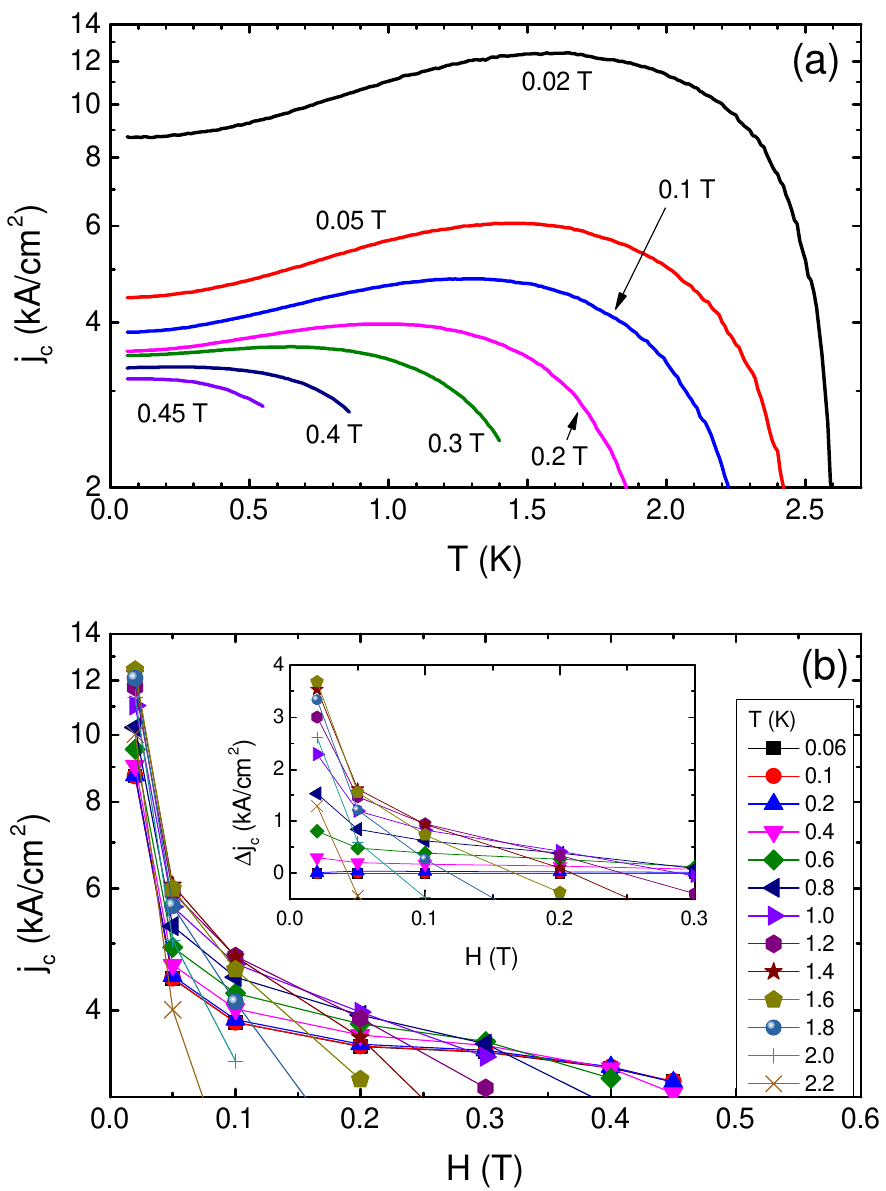}%
\caption{\label{fig:jc} Calculated theoretical critical current (a) as a function of temperature in different magnetic fields and (b) as a function of magnetic field at fixed temperatures. Inset in (b) displays $\Delta j_c(T,H)=j_c(T,H)-j_c(0.06 ~\textmd{K}, H)$ to show the maximum $j_c$ being at a intermediate temperature.}
\end{figure}

In the approximation of a linear elastic response of a vortex lattice to a small - amplitude AC perturbation (which is the case here), the total magnetic penetration depth can be expressed as a sum of Meissner and vortex contributions, which are represented by the London and the Campbell penetration depths, respectively, $\lambda_m^2=\lambda^2 +\lambda_C^2$. \cite{Campbell1969,Campbell1971,Brandt1995} Since we know $\lambda$ from the measurements in zero field, we can readily calculate $\lambda_C$. The calculated Campbell penetration depth as a function of magnetic field shown in Fig.~\ref{fig:Campbell} appears sub-linear as expected in conventional superconductors ($\sim \sqrt{H}$) at a first glance. However, non-monotonic behavior is revealed upon closer inspection as shown in the inset. This non-monotonic behavior at low fields originates from non-monotonic temperature dependence discussed above.

\subsection{Theoretical critical current}

\begin{figure}[tbh]
\includegraphics[width=1.0\linewidth]{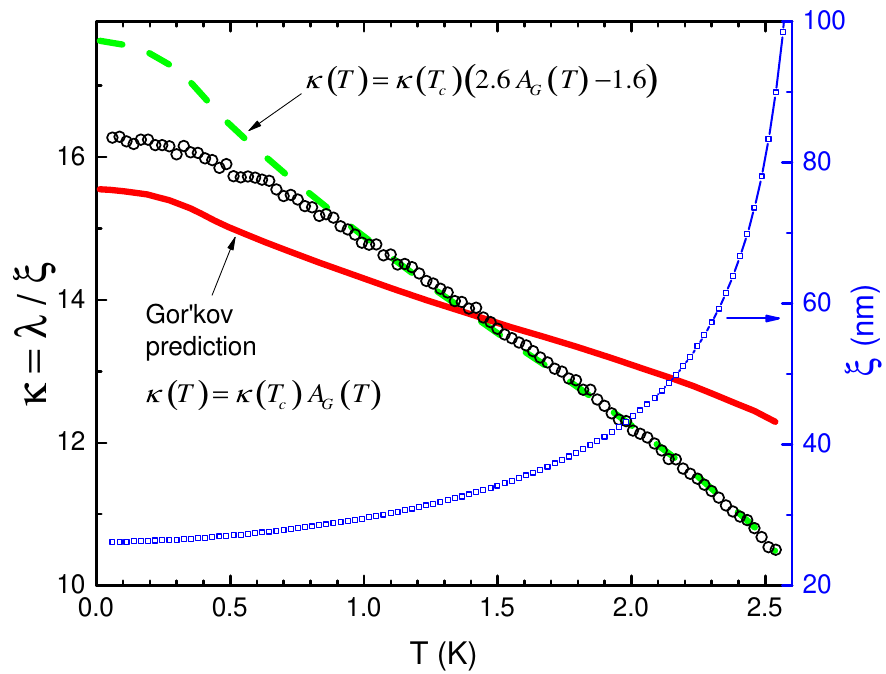}%
\caption{\label{fig:kappa} Calculated Ginzburg - Landau parameter, $\kappa(T) = \lambda(T)/\xi(T)$ (symbols, left axis) compared to the Gor'kov prediction, $\kappa \left( T \right) = \kappa \left( T_c \right)A_G\left( T \right)$ (solid line) and to modified to best fit the data expression, $\kappa \left( T \right) = \kappa \left( T_c \right)\left( 2.6A_G\left( T \right) - 1.6 \right)$ (dashed line). Right axis shows estimated coherence length.}
\end{figure}

\begin{figure}[tb]
\includegraphics[width=1.0\linewidth]{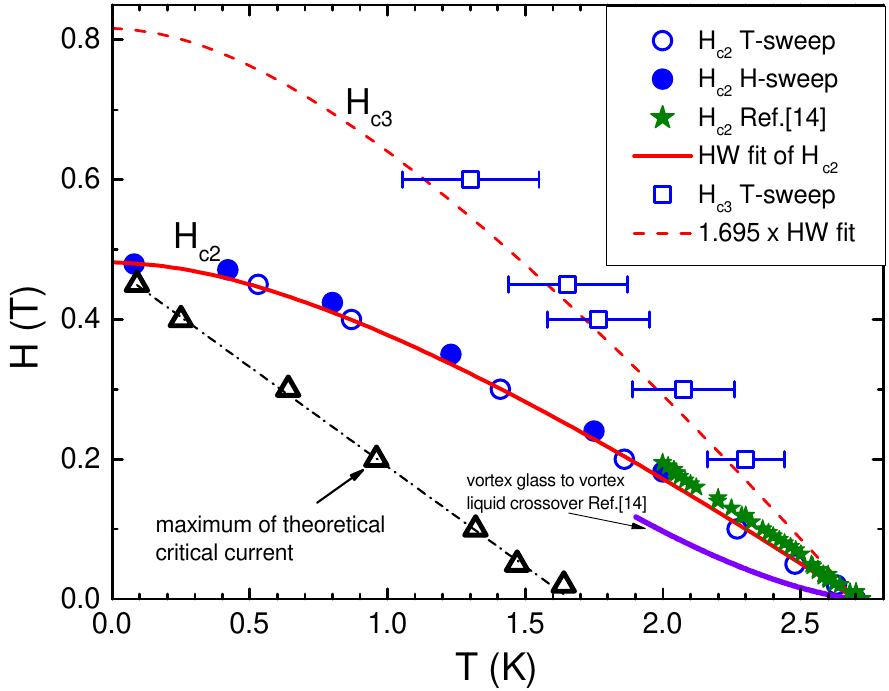}%
\caption{\label{fig:HT} $H$-$T$ phase diagram. Open and solid circles represent bulk superconducting transition, $H_{c2}(T)$, determined by temperature- and field-sweepings, respectively. Red solid line is fitting over experimental $H_{c2}(T)$ using the HW theory. Open squares indicate the temperatures where diamagnetic response was detected above bulk $T_c$. Red dashed line represents $H_{c3}=1.695H_{c2}(T)$. Open triangles represent the strongest pinning. Black dashed dots are a linear line fitting for the strongest pinning. For comparison, we include the data for $H_{c2}$ (green stars) and a crossover from vortex liquid to vortex glass phase (violet solid line) from Ref.~\onlinecite{Sung2012a}.}
\end{figure}

We distinguish theoretical critical current, which is a parameter entering the simple expression for the Campbell penetration depth, $\frac{4\pi}{c}j_c=r_p \phi_0/\lambda_C^2$ from the actual critical current that is affected by intervortex interactions and from the measured critical (persistent) current that is further affected by magnetic relaxation. Here $r_p$ is a characteristic radius of the pinning potential. We assume the $r_p$ to be equal to the coherence length $\xi(T) = \sqrt{\phi_0/2\pi H_{c2}(T)}$. Obtained theoretical critical current density is shown in Fig.~\ref{fig:jc}, (a) as function of temperature at different fields and (b) - as a function of an applied field at different temperatures. The critical current is non - monotonic as a function of temperature showing maximum at the intermediate temperatures. We attribute this to the matching between temperature - dependent coherence length and pinning landscape. This assertion is plausible given monotonic behavior of the critical current versus field where the variation of the coherence length is much weaker.\cite{Kogan2006} Comparison with the previous works also re-affirms that we are dealing with the upper theoretical estimate of the critical current, approximately four times larger than estimated from the magnetic measurements.\cite{Sung2012a} On the other hand, direct comparison with our data shown in Fig.~\ref{fig:HT} shows that the maximum current density line found in this work is not an extension of the irreversibility line found in Ref.~\onlinecite{Sung2012a} and probably represents a crossover in the pinning mechanism reflected in the change of the effective pinning potential.

Unfortunately literature data are only available for higher temperatures above $T=2$ K and weak fields less than $H=0.1$ T, so we do not know whether the nonmonotonic behavior of $j_c(T)$ propagates to the relaxed persistent current density.

Furthermore, our evaluation of the coherence length, $\xi(T)$ (right axis in Fig.~\ref{fig:kappa}), allows experimental determination of the temperature - dependent Ginzburg - Landau parameter, $\kappa(T) = \lambda(T)/\xi(T)$, shown in Fig.~\ref{fig:kappa} (left axis). The result is compared with the Gor'kov theory \cite{Gorkov60} where temperature correction is introduced as $A_G(T)=\kappa(T)/\kappa(T_c)$. The trend is correct, but the magnitude of the variation is smaller then the observed. If we attempt to rescale the Gor'kov's result, the best agreement at the intermediate temperatures is achieved with $\kappa \left( T \right) = \kappa \left( T_c \right)\left( 2.6A_G\left( T \right) - 1.6 \right)$. Perhaps a better agreement could be achieved with a realistic band structure.

\subsection{$H-T$ phase diagram}

Finally, a $H$-$T$ phase diagram is established based on the measured $\Delta\lambda_m(T,H)$ as shown in Fig.~\ref{fig:HT}. Open and solid circles represent bulk superconducting phase transitions determined by temperature- and field- sweep measurements, respectively. The red solid line throughout these symbols is the HW - fit \cite{Helfand1966} with $H_{c2}(0)=0.4817$~T. This value was determined with the experimental data down to $0.02T_c$ in our study, and it is very close to the estimated value $H_{c2}(0)=0.4920$ T in Ref.~\onlinecite{Sung2011} only with the the slope at $T_c$ by using the same theory. This is a good indication that $H_{c2}(T)$ of SrPd$_2$Ge$_2$ is orbit - limited.

The additional diamagnetic phase between bulk $T_c$ and normal state detected in temperature sweeps is marked by open squares. This phase transition is clearest in magnetic fields between 0.2 T and 0.6 T. A linear fitting over these squares extrapolates down to almost at $H_{c2}(T_c)$ and up to 1 T where the diamagnetic phase was not seen.
It is well known that surface superconductivity boundary is given by $H_{c3}(T) = 1.695 H_{c2}(T)$.\cite{Tinkham2004} This curve fits well the diamagnetic feature observed in our experiments, so it makes sense to assign it to surface superconductivity.

Lastly, the strongest pinning region where theoretical critical current is maximum is marked by open triangles with a linear fitting line shown in dashed dots. It lies deep inside the superconducting state and is probably due to the matching effect. For comparison, we include the data for $H_{c2}$ (green stars) and a crossover from vortex liquid to vortex glass phase (violet solid line), both estimated from the magnetization measurements.\cite{Sung2012a} The upper critical field is in a good agreement with our data. The liquid-glass crossover line, on the other hand, does not extrapolate to the line of the strong pinning found in our measurements (although there is no overlapping temperature interval). This indicates that the maximum theoretical critical current line is not an extension of the irreversibility line and represents a different crossover in the pinning mechanism reflected in the non-monotonic change in the shape of the pinning potential.

\section{Summary}
Temperature-and magnetic field-dependent penetration depth was measured in single crystals of SrPd$_2$Ge$_2$ superconductor by using a tunnel diode resonator technique. For $H=0$, the London penetration depth saturates at low temperatures indicating a fully gapped superconductivity in SrPd$_2$Ge$_2$. The calculated superfluid density is best described by a single - gap s-wave superconductor, perhaps slightly on the stronger coupling side. The $H_{c2}(T)$ was measured down to $0.02T_c$ and could be well described by the HW theory with $H_{c2}(0)= 0.4817$~T and is clearly limited by the orbital depairing. In the magnetic fields between 0.2 T and 0.6 T, another diamagnetic phase other than bulk superconductivity is clearly seen, and this phase line is consistent with the surface superconductivity bound by the third critical field, $H_{c3}=1.695H_{c2}$. The Campbell penetration depth and the theoretical critical current density both exhibit non-monototic behavior with the strongest pinning at intermediate temperatures, probably due to matching effects between the temperature - dependent coherence length and the relevant pinning landscape lengthscale.

\begin{acknowledgments}
RP, MAT and HK thank V. Kogan for insightful comments. HK and NHS thank Jeehoon Kim for useful discussions. The work at Ames was supported by the U.S. Department of Energy, Office of Basic Energy Sciences, Division of Materials Sciences and Engineering under contract No. DE-AC02-07CH11358. The work at GIST was supported by the Ministry of Education, Science and Technology of the Republic of Korea (2011-0028736).
\end{acknowledgments}


\begin{thebibliography}{32}%
\makeatletter
\providecommand \@ifxundefined [1]{%
 \@ifx{#1\undefined}
}%
\providecommand \@ifnum [1]{%
 \ifnum #1\expandafter \@firstoftwo
 \else \expandafter \@secondoftwo
 \fi
}%
\providecommand \@ifx [1]{%
 \ifx #1\expandafter \@firstoftwo
 \else \expandafter \@secondoftwo
 \fi
}%
\providecommand \natexlab [1]{#1}%
\providecommand \enquote  [1]{``#1''}%
\providecommand \bibnamefont  [1]{#1}%
\providecommand \bibfnamefont [1]{#1}%
\providecommand \citenamefont [1]{#1}%
\providecommand \href@noop [0]{\@secondoftwo}%
\providecommand \href [0]{\begingroup \@sanitize@url \@href}%
\providecommand \@href[1]{\@@startlink{#1}\@@href}%
\providecommand \@@href[1]{\endgroup#1\@@endlink}%
\providecommand \@sanitize@url [0]{\catcode `\\12\catcode `\$12\catcode
  `\&12\catcode `\#12\catcode `\^12\catcode `\_12\catcode `\%12\relax}%
\providecommand \@@startlink[1]{}%
\providecommand \@@endlink[0]{}%
\providecommand \url  [0]{\begingroup\@sanitize@url \@url }%
\providecommand \@url [1]{\endgroup\@href {#1}{\urlprefix }}%
\providecommand \urlprefix  [0]{URL }%
\providecommand \Eprint [0]{\href }%
\providecommand \doibase [0]{http://dx.doi.org/}%
\providecommand \selectlanguage [0]{\@gobble}%
\providecommand \bibinfo  [0]{\@secondoftwo}%
\providecommand \bibfield  [0]{\@secondoftwo}%
\providecommand \translation [1]{[#1]}%
\providecommand \BibitemOpen [0]{}%
\providecommand \bibitemStop [0]{}%
\providecommand \bibitemNoStop [0]{.\EOS\space}%
\providecommand \EOS [0]{\spacefactor3000\relax}%
\providecommand \BibitemShut  [1]{\csname bibitem#1\endcsname}%
\let\auto@bib@innerbib\@empty
\bibitem [{\citenamefont {Fujii}\ and\ \citenamefont {Sato}(2009)}]{Fujii2009}%
  \BibitemOpen
  \bibfield  {author} {\bibinfo {author} {\bibfnamefont {H.}~\bibnamefont
  {Fujii}}\ and\ \bibinfo {author} {\bibfnamefont {A.}~\bibnamefont {Sato}},\
  }\href {\doibase 10.1016/j.jallcom.2009.08.050} {\bibfield  {journal}
  {\bibinfo  {journal} {Journal of Alloys and Compounds}\ }\textbf {\bibinfo
  {volume} {487}},\ \bibinfo {pages} {198 } (\bibinfo {year}
  {2009})}\BibitemShut {NoStop}%
\bibitem [{\citenamefont {Sung}\ \emph {et~al.}(2011)\citenamefont {Sung},
  \citenamefont {Rhyee},\ and\ \citenamefont {Cho}}]{Sung2011}%
  \BibitemOpen
  \bibfield  {author} {\bibinfo {author} {\bibfnamefont {N.~H.}\ \bibnamefont
  {Sung}}, \bibinfo {author} {\bibfnamefont {J.-S.}\ \bibnamefont {Rhyee}}, \
  and\ \bibinfo {author} {\bibfnamefont {B.~K.}\ \bibnamefont {Cho}},\ }\href
  {\doibase 10.1103/PhysRevB.83.094511} {\bibfield  {journal} {\bibinfo
  {journal} {Phys. Rev. B}\ }\textbf {\bibinfo {volume} {83}},\ \bibinfo
  {pages} {094511} (\bibinfo {year} {2011})}\BibitemShut {NoStop}%
\bibitem [{\citenamefont {Helfand}\ and\ \citenamefont
  {Werthamer}(1966)}]{Helfand1966}%
  \BibitemOpen
  \bibfield  {author} {\bibinfo {author} {\bibfnamefont {E.}~\bibnamefont
  {Helfand}}\ and\ \bibinfo {author} {\bibfnamefont {N.~R.}\ \bibnamefont
  {Werthamer}},\ }\href {\doibase 10.1103/PhysRev.147.288} {\bibfield
  {journal} {\bibinfo  {journal} {Phys. Rev.}\ }\textbf {\bibinfo {volume}
  {147}},\ \bibinfo {pages} {288} (\bibinfo {year} {1966})}\BibitemShut
  {NoStop}%
\bibitem [{\citenamefont {Sung}\ \emph
  {et~al.}(2012{\natexlab{a}})\citenamefont {Sung}, \citenamefont {Roh},
  \citenamefont {Kang},\ and\ \citenamefont {Cho}}]{Sung2012}%
  \BibitemOpen
  \bibfield  {author} {\bibinfo {author} {\bibfnamefont {N.~H.}\ \bibnamefont
  {Sung}}, \bibinfo {author} {\bibfnamefont {C.~J.}\ \bibnamefont {Roh}},
  \bibinfo {author} {\bibfnamefont {B.~Y.}\ \bibnamefont {Kang}}, \ and\
  \bibinfo {author} {\bibfnamefont {B.~K.}\ \bibnamefont {Cho}},\ }\href
  {\doibase 10.1063/1.3672089} {\bibfield  {journal} {\bibinfo  {journal}
  {Journal of Applied Physics}\ }\textbf {\bibinfo {volume} {111}},\ \bibinfo
  {eid} {07E117} (\bibinfo {year} {2012}{\natexlab{a}})}\BibitemShut {NoStop}%
\bibitem [{\citenamefont {Kim}\ \emph {et~al.}(2012{\natexlab{a}})\citenamefont
  {Kim}, \citenamefont {Yaresko}, \citenamefont {Zabolotnyy}, \citenamefont
  {Kordyuk}, \citenamefont {Evtushinsky}, \citenamefont {Sung}, \citenamefont
  {Cho}, \citenamefont {Samuely}, \citenamefont {Szab\'o}, \citenamefont
  {Rodrigo}, \citenamefont {Park}, \citenamefont {Inosov}, \citenamefont
  {Samuely}, \citenamefont {B\"uchner},\ and\ \citenamefont
  {Borisenko}}]{Kim2012}%
  \BibitemOpen
  \bibfield  {author} {\bibinfo {author} {\bibfnamefont {T.~K.}\ \bibnamefont
  {Kim}}, \bibinfo {author} {\bibfnamefont {A.~N.}\ \bibnamefont {Yaresko}},
  \bibinfo {author} {\bibfnamefont {V.~B.}\ \bibnamefont {Zabolotnyy}},
  \bibinfo {author} {\bibfnamefont {A.~A.}\ \bibnamefont {Kordyuk}}, \bibinfo
  {author} {\bibfnamefont {D.~V.}\ \bibnamefont {Evtushinsky}}, \bibinfo
  {author} {\bibfnamefont {N.~H.}\ \bibnamefont {Sung}}, \bibinfo {author}
  {\bibfnamefont {B.~K.}\ \bibnamefont {Cho}}, \bibinfo {author} {\bibfnamefont
  {T.}~\bibnamefont {Samuely}}, \bibinfo {author} {\bibfnamefont
  {P.}~\bibnamefont {Szab\'o}}, \bibinfo {author} {\bibfnamefont {J.~G.}\
  \bibnamefont {Rodrigo}}, \bibinfo {author} {\bibfnamefont {J.~T.}\
  \bibnamefont {Park}}, \bibinfo {author} {\bibfnamefont {D.~S.}\ \bibnamefont
  {Inosov}}, \bibinfo {author} {\bibfnamefont {P.}~\bibnamefont {Samuely}},
  \bibinfo {author} {\bibfnamefont {B.}~\bibnamefont {B\"uchner}}, \ and\
  \bibinfo {author} {\bibfnamefont {S.~V.}\ \bibnamefont {Borisenko}},\ }\href
  {\doibase 10.1103/PhysRevB.85.014520} {\bibfield  {journal} {\bibinfo
  {journal} {Phys. Rev. B}\ }\textbf {\bibinfo {volume} {85}},\ \bibinfo
  {pages} {014520} (\bibinfo {year} {2012}{\natexlab{a}})}\BibitemShut
  {NoStop}%
\bibitem [{\citenamefont {Samuely}\ \emph {et~al.}(2013)\citenamefont
  {Samuely}, \citenamefont {Szab\'{o}}, \citenamefont {Pribulov\'{o}},
  \citenamefont {Sung}, \citenamefont {Cho}, \citenamefont {Klein},
  \citenamefont {Cambel}, \citenamefont {Rodrigo},\ and\ \citenamefont
  {Samuely}}]{Samuely2013}%
  \BibitemOpen
  \bibfield  {author} {\bibinfo {author} {\bibfnamefont {T.}~\bibnamefont
  {Samuely}}, \bibinfo {author} {\bibfnamefont {P.}~\bibnamefont {Szab\'{o}}},
  \bibinfo {author} {\bibfnamefont {Z.}~\bibnamefont {Pribulov\'{o}}}, \bibinfo
  {author} {\bibfnamefont {N.~H.}\ \bibnamefont {Sung}}, \bibinfo {author}
  {\bibfnamefont {B.~K.}\ \bibnamefont {Cho}}, \bibinfo {author} {\bibfnamefont
  {T.}~\bibnamefont {Klein}}, \bibinfo {author} {\bibfnamefont
  {V.}~\bibnamefont {Cambel}}, \bibinfo {author} {\bibfnamefont {J.~G.}\
  \bibnamefont {Rodrigo}}, \ and\ \bibinfo {author} {\bibfnamefont
  {P.}~\bibnamefont {Samuely}},\ }\href
  {http://stacks.iop.org/0953-2048/26/i=1/a=015010} {\bibfield  {journal}
  {\bibinfo  {journal} {Superconductor Science and Technology}\ }\textbf
  {\bibinfo {volume} {26}},\ \bibinfo {pages} {015010} (\bibinfo {year}
  {2013})}\BibitemShut {NoStop}%
\bibitem [{\citenamefont {Hashimoto}\ \emph {et~al.}(2010)\citenamefont
  {Hashimoto}, \citenamefont {Serafin}, \citenamefont {Tonegawa}, \citenamefont
  {Katsumata}, \citenamefont {Okazaki}, \citenamefont {Saito}, \citenamefont
  {Fukazawa}, \citenamefont {Kohori}, \citenamefont {Kihou}, \citenamefont
  {Lee}, \citenamefont {Iyo}, \citenamefont {Eisaki}, \citenamefont {Ikeda},
  \citenamefont {Matsuda}, \citenamefont {Carrington},\ and\ \citenamefont
  {Shibauchi}}]{Hashimoto2010}%
  \BibitemOpen
  \bibfield  {author} {\bibinfo {author} {\bibfnamefont {K.}~\bibnamefont
  {Hashimoto}}, \bibinfo {author} {\bibfnamefont {A.}~\bibnamefont {Serafin}},
  \bibinfo {author} {\bibfnamefont {S.}~\bibnamefont {Tonegawa}}, \bibinfo
  {author} {\bibfnamefont {R.}~\bibnamefont {Katsumata}}, \bibinfo {author}
  {\bibfnamefont {R.}~\bibnamefont {Okazaki}}, \bibinfo {author} {\bibfnamefont
  {T.}~\bibnamefont {Saito}}, \bibinfo {author} {\bibfnamefont
  {H.}~\bibnamefont {Fukazawa}}, \bibinfo {author} {\bibfnamefont
  {Y.}~\bibnamefont {Kohori}}, \bibinfo {author} {\bibfnamefont
  {K.}~\bibnamefont {Kihou}}, \bibinfo {author} {\bibfnamefont {C.~H.}\
  \bibnamefont {Lee}}, \bibinfo {author} {\bibfnamefont {A.}~\bibnamefont
  {Iyo}}, \bibinfo {author} {\bibfnamefont {H.}~\bibnamefont {Eisaki}},
  \bibinfo {author} {\bibfnamefont {H.}~\bibnamefont {Ikeda}}, \bibinfo
  {author} {\bibfnamefont {Y.}~\bibnamefont {Matsuda}}, \bibinfo {author}
  {\bibfnamefont {A.}~\bibnamefont {Carrington}}, \ and\ \bibinfo {author}
  {\bibfnamefont {T.}~\bibnamefont {Shibauchi}},\ }\href {\doibase
  10.1103/PhysRevB.82.014526} {\bibfield  {journal} {\bibinfo  {journal} {Phys.
  Rev. B}\ }\textbf {\bibinfo {volume} {82}},\ \bibinfo {pages} {014526}
  (\bibinfo {year} {2010})}\BibitemShut {NoStop}%
\bibitem [{\citenamefont {Reid}\ \emph {et~al.}(2012)\citenamefont {Reid},
  \citenamefont {Tanatar}, \citenamefont {Juneau-Fecteau}, \citenamefont
  {Gordon}, \citenamefont {de~Cotret}, \citenamefont {Doiron-Leyraud},
  \citenamefont {Saito}, \citenamefont {Fukazawa}, \citenamefont {Kohori},
  \citenamefont {Kihou}, \citenamefont {Lee}, \citenamefont {Iyo},
  \citenamefont {Eisaki}, \citenamefont {Prozorov},\ and\ \citenamefont
  {Taillefer}}]{Reid2012}%
  \BibitemOpen
  \bibfield  {author} {\bibinfo {author} {\bibfnamefont {J.-P.}\ \bibnamefont
  {Reid}}, \bibinfo {author} {\bibfnamefont {M.~A.}\ \bibnamefont {Tanatar}},
  \bibinfo {author} {\bibfnamefont {A.}~\bibnamefont {Juneau-Fecteau}},
  \bibinfo {author} {\bibfnamefont {R.~T.}\ \bibnamefont {Gordon}}, \bibinfo
  {author} {\bibfnamefont {S.~R.}\ \bibnamefont {de~Cotret}}, \bibinfo {author}
  {\bibfnamefont {N.}~\bibnamefont {Doiron-Leyraud}}, \bibinfo {author}
  {\bibfnamefont {T.}~\bibnamefont {Saito}}, \bibinfo {author} {\bibfnamefont
  {H.}~\bibnamefont {Fukazawa}}, \bibinfo {author} {\bibfnamefont
  {Y.}~\bibnamefont {Kohori}}, \bibinfo {author} {\bibfnamefont
  {K.}~\bibnamefont {Kihou}}, \bibinfo {author} {\bibfnamefont {C.~H.}\
  \bibnamefont {Lee}}, \bibinfo {author} {\bibfnamefont {A.}~\bibnamefont
  {Iyo}}, \bibinfo {author} {\bibfnamefont {H.}~\bibnamefont {Eisaki}},
  \bibinfo {author} {\bibfnamefont {R.}~\bibnamefont {Prozorov}}, \ and\
  \bibinfo {author} {\bibfnamefont {L.}~\bibnamefont {Taillefer}},\ }\href
  {\doibase 10.1103/PhysRevLett.109.087001} {\bibfield  {journal} {\bibinfo
  {journal} {Phys. Rev. Lett.}\ }\textbf {\bibinfo {volume} {109}},\ \bibinfo
  {pages} {087001} (\bibinfo {year} {2012})}\BibitemShut {NoStop}%
\bibitem [{\citenamefont {Kurita}\ \emph {et~al.}(2009)\citenamefont {Kurita},
  \citenamefont {Ronning}, \citenamefont {Tokiwa}, \citenamefont {Bauer},
  \citenamefont {Subedi}, \citenamefont {Singh}, \citenamefont {Thompson},\
  and\ \citenamefont {Movshovich}}]{Kurita2009}%
  \BibitemOpen
  \bibfield  {author} {\bibinfo {author} {\bibfnamefont {N.}~\bibnamefont
  {Kurita}}, \bibinfo {author} {\bibfnamefont {F.}~\bibnamefont {Ronning}},
  \bibinfo {author} {\bibfnamefont {Y.}~\bibnamefont {Tokiwa}}, \bibinfo
  {author} {\bibfnamefont {E.~D.}\ \bibnamefont {Bauer}}, \bibinfo {author}
  {\bibfnamefont {A.}~\bibnamefont {Subedi}}, \bibinfo {author} {\bibfnamefont
  {D.~J.}\ \bibnamefont {Singh}}, \bibinfo {author} {\bibfnamefont {J.~D.}\
  \bibnamefont {Thompson}}, \ and\ \bibinfo {author} {\bibfnamefont
  {R.}~\bibnamefont {Movshovich}},\ }\href {\doibase
  10.1103/PhysRevLett.102.147004} {\bibfield  {journal} {\bibinfo  {journal}
  {Phys. Rev. Lett.}\ }\textbf {\bibinfo {volume} {102}},\ \bibinfo {pages}
  {147004} (\bibinfo {year} {2009})}\BibitemShut {NoStop}%
\bibitem [{\citenamefont {Kurita}\ \emph {et~al.}(2011)\citenamefont {Kurita},
  \citenamefont {Ronning}, \citenamefont {Miclea}, \citenamefont {Bauer},
  \citenamefont {Gofryk}, \citenamefont {Thompson},\ and\ \citenamefont
  {Movshovich}}]{Kurita2011}%
  \BibitemOpen
  \bibfield  {author} {\bibinfo {author} {\bibfnamefont {N.}~\bibnamefont
  {Kurita}}, \bibinfo {author} {\bibfnamefont {F.}~\bibnamefont {Ronning}},
  \bibinfo {author} {\bibfnamefont {C.~F.}\ \bibnamefont {Miclea}}, \bibinfo
  {author} {\bibfnamefont {E.~D.}\ \bibnamefont {Bauer}}, \bibinfo {author}
  {\bibfnamefont {K.}~\bibnamefont {Gofryk}}, \bibinfo {author} {\bibfnamefont
  {J.~D.}\ \bibnamefont {Thompson}}, \ and\ \bibinfo {author} {\bibfnamefont
  {R.}~\bibnamefont {Movshovich}},\ }\href {\doibase
  10.1103/PhysRevB.83.094527} {\bibfield  {journal} {\bibinfo  {journal} {Phys.
  Rev. B}\ }\textbf {\bibinfo {volume} {83}},\ \bibinfo {pages} {094527}
  (\bibinfo {year} {2011})}\BibitemShut {NoStop}%
\bibitem [{\citenamefont {Prozorov}\ and\ \citenamefont
  {Giannetta}(2006)}]{Prozorov2006}%
  \BibitemOpen
  \bibfield  {author} {\bibinfo {author} {\bibfnamefont {R.}~\bibnamefont
  {Prozorov}}\ and\ \bibinfo {author} {\bibfnamefont {R.~W.}\ \bibnamefont
  {Giannetta}},\ }\href {http://stacks.iop.org/0953-2048/19/i=8/a=R01}
  {\bibfield  {journal} {\bibinfo  {journal} {Superconductor Science and
  Technology}\ }\textbf {\bibinfo {volume} {19}},\ \bibinfo {pages} {R41}
  (\bibinfo {year} {2006})}\BibitemShut {NoStop}%
\bibitem [{\citenamefont {Prozorov}\ and\ \citenamefont
  {Kogan}(2011)}]{ProzorovKoganROPP2011}%
  \BibitemOpen
  \bibfield  {author} {\bibinfo {author} {\bibfnamefont {R.}~\bibnamefont
  {Prozorov}}\ and\ \bibinfo {author} {\bibfnamefont {V.~G.}\ \bibnamefont
  {Kogan}},\ }\href {\doibase 10.1088/0034-4885/74/12/124505} {\bibfield
  {journal} {\bibinfo  {journal} {Reports on Progress in Physics}\ }\textbf
  {\bibinfo {volume} {74}},\ \bibinfo {pages} {124505} (\bibinfo {year}
  {2011})}\BibitemShut {NoStop}%
\bibitem [{\citenamefont {Brandt}(1995)}]{Brandt1995}%
  \BibitemOpen
  \bibfield  {author} {\bibinfo {author} {\bibfnamefont {E.~H.}\ \bibnamefont
  {Brandt}},\ }\href {http://stacks.iop.org/0034-4885/58/1465} {\bibfield
  {journal} {\bibinfo  {journal} {Reports on Progress in Physics}\ }\textbf
  {\bibinfo {volume} {58}},\ \bibinfo {pages} {1465} (\bibinfo {year}
  {1995})}\BibitemShut {NoStop}%
\bibitem [{\citenamefont {Sung}\ \emph
  {et~al.}(2012{\natexlab{b}})\citenamefont {Sung}, \citenamefont {Jo},\ and\
  \citenamefont {Cho}}]{Sung2012a}%
  \BibitemOpen
  \bibfield  {author} {\bibinfo {author} {\bibfnamefont {N.~H.}\ \bibnamefont
  {Sung}}, \bibinfo {author} {\bibfnamefont {Y.~J.}\ \bibnamefont {Jo}}, \ and\
  \bibinfo {author} {\bibfnamefont {B.~K.}\ \bibnamefont {Cho}},\ }\href
  {http://stacks.iop.org/0953-2048/25/i=7/a=075002} {\bibfield  {journal}
  {\bibinfo  {journal} {Superconductor Science and Technology}\ }\textbf
  {\bibinfo {volume} {25}},\ \bibinfo {pages} {075002} (\bibinfo {year}
  {2012}{\natexlab{b}})}\BibitemShut {NoStop}%
\bibitem [{\citenamefont {Tinkham}(2004)}]{Tinkham2004}%
  \BibitemOpen
  \bibfield  {author} {\bibinfo {author} {\bibfnamefont {M.}~\bibnamefont
  {Tinkham}},\ }\href {http://www.worldcat.org/isbn/0486435032} {\emph
  {\bibinfo {title} {Introduction to Superconductivity: Second Edition}}},\
  \bibinfo {edition} {2nd}\ ed.\ (\bibinfo  {publisher} {Dover Publications},\
  \bibinfo {year} {2004})\BibitemShut {NoStop}%
\bibitem [{\citenamefont {Prozorov}\ \emph {et~al.}(2000)\citenamefont
  {Prozorov}, \citenamefont {Giannetta}, \citenamefont {Carrington},\ and\
  \citenamefont {Araujo-Moreira}}]{Prozorov2000}%
  \BibitemOpen
  \bibfield  {author} {\bibinfo {author} {\bibfnamefont {R.}~\bibnamefont
  {Prozorov}}, \bibinfo {author} {\bibfnamefont {R.~W.}\ \bibnamefont
  {Giannetta}}, \bibinfo {author} {\bibfnamefont {A.}~\bibnamefont
  {Carrington}}, \ and\ \bibinfo {author} {\bibfnamefont {F.~M.}\ \bibnamefont
  {Araujo-Moreira}},\ }\href {\doibase 10.1103/PhysRevB.62.115} {\bibfield
  {journal} {\bibinfo  {journal} {Phys. Rev. B}\ }\textbf {\bibinfo {volume}
  {62}},\ \bibinfo {pages} {115} (\bibinfo {year} {2000})}\BibitemShut
  {NoStop}%
\bibitem [{\citenamefont {Fletcher}\ \emph {et~al.}(2005)\citenamefont
  {Fletcher}, \citenamefont {Carrington}, \citenamefont {Taylor}, \citenamefont
  {Kazakov},\ and\ \citenamefont {Karpinski}}]{FletcherMgB2}%
  \BibitemOpen
  \bibfield  {author} {\bibinfo {author} {\bibfnamefont {J.~D.}\ \bibnamefont
  {Fletcher}}, \bibinfo {author} {\bibfnamefont {A.}~\bibnamefont
  {Carrington}}, \bibinfo {author} {\bibfnamefont {O.~J.}\ \bibnamefont
  {Taylor}}, \bibinfo {author} {\bibfnamefont {S.~M.}\ \bibnamefont {Kazakov}},
  \ and\ \bibinfo {author} {\bibfnamefont {J.}~\bibnamefont {Karpinski}},\
  }\href {\doibase 10.1103/PhysRevLett.95.097005} {\bibfield  {journal}
  {\bibinfo  {journal} {Phys. Rev. Lett.}\ }\textbf {\bibinfo {volume} {95}},\
  \bibinfo {pages} {097005} (\bibinfo {year} {2005})}\BibitemShut {NoStop}%
\bibitem [{\citenamefont {Fletcher}\ \emph {et~al.}(2007)\citenamefont
  {Fletcher}, \citenamefont {Carrington}, \citenamefont {Diener}, \citenamefont
  {Rodiere}, \citenamefont {Brison}, \citenamefont {Prozorov}, \citenamefont
  {Olheiser},\ and\ \citenamefont {Giannetta}}]{Fletcher2007NbSe2}%
  \BibitemOpen
  \bibfield  {author} {\bibinfo {author} {\bibfnamefont {J.~D.}\ \bibnamefont
  {Fletcher}}, \bibinfo {author} {\bibfnamefont {A.}~\bibnamefont
  {Carrington}}, \bibinfo {author} {\bibfnamefont {P.}~\bibnamefont {Diener}},
  \bibinfo {author} {\bibfnamefont {P.}~\bibnamefont {Rodiere}}, \bibinfo
  {author} {\bibfnamefont {J.~P.}\ \bibnamefont {Brison}}, \bibinfo {author}
  {\bibfnamefont {R.}~\bibnamefont {Prozorov}}, \bibinfo {author}
  {\bibfnamefont {T.}~\bibnamefont {Olheiser}}, \ and\ \bibinfo {author}
  {\bibfnamefont {R.~W.}\ \bibnamefont {Giannetta}},\ }\href@noop {} {\bibfield
   {journal} {\bibinfo  {journal} {Phys. Rev. Lett.}\ }\textbf {\bibinfo
  {volume} {98}},\ \bibinfo {pages} {057003} (\bibinfo {year}
  {2007})}\BibitemShut {NoStop}%
\bibitem [{\citenamefont {Gordon}\ \emph {et~al.}(2008)\citenamefont {Gordon},
  \citenamefont {Vannette}, \citenamefont {Martin}, \citenamefont {Nakajima},
  \citenamefont {Tamegai},\ and\ \citenamefont {Prozorov}}]{Gordon2008}%
  \BibitemOpen
  \bibfield  {author} {\bibinfo {author} {\bibfnamefont {R.~T.}\ \bibnamefont
  {Gordon}}, \bibinfo {author} {\bibfnamefont {M.~D.}\ \bibnamefont
  {Vannette}}, \bibinfo {author} {\bibfnamefont {C.}~\bibnamefont {Martin}},
  \bibinfo {author} {\bibfnamefont {Y.}~\bibnamefont {Nakajima}}, \bibinfo
  {author} {\bibfnamefont {T.}~\bibnamefont {Tamegai}}, \ and\ \bibinfo
  {author} {\bibfnamefont {R.}~\bibnamefont {Prozorov}},\ }\href@noop {}
  {\bibfield  {journal} {\bibinfo  {journal} {Phys. Rev. B}\ }\textbf {\bibinfo
  {volume} {78}},\ \bibinfo {pages} {024514} (\bibinfo {year}
  {2008})}\BibitemShut {NoStop}%
\bibitem [{\citenamefont {Kim}\ \emph {et~al.}(2011)\citenamefont {Kim},
  \citenamefont {Tanatar}, \citenamefont {Song}, \citenamefont {Kwon},\ and\
  \citenamefont {Prozorov}}]{Kim2011}%
  \BibitemOpen
  \bibfield  {author} {\bibinfo {author} {\bibfnamefont {H.}~\bibnamefont
  {Kim}}, \bibinfo {author} {\bibfnamefont {M.~A.}\ \bibnamefont {Tanatar}},
  \bibinfo {author} {\bibfnamefont {Y.~J.}\ \bibnamefont {Song}}, \bibinfo
  {author} {\bibfnamefont {Y.~S.}\ \bibnamefont {Kwon}}, \ and\ \bibinfo
  {author} {\bibfnamefont {R.}~\bibnamefont {Prozorov}},\ }\href {\doibase
  10.1103/PhysRevB.83.100502} {\bibfield  {journal} {\bibinfo  {journal} {Phys.
  Rev. B}\ }\textbf {\bibinfo {volume} {83}},\ \bibinfo {pages} {100502}
  (\bibinfo {year} {2011})}\BibitemShut {NoStop}%
\bibitem [{\citenamefont {Lifshitz}\ \emph {et~al.}(1984)\citenamefont
  {Lifshitz}, \citenamefont {Landau},\ and\ \citenamefont
  {Pitaevskii}}]{Lifshitz1984}%
  \BibitemOpen
  \bibfield  {author} {\bibinfo {author} {\bibfnamefont {E.~M.}\ \bibnamefont
  {Lifshitz}}, \bibinfo {author} {\bibfnamefont {L.~D.}\ \bibnamefont
  {Landau}}, \ and\ \bibinfo {author} {\bibfnamefont {L.~P.}\ \bibnamefont
  {Pitaevskii}},\ }\href {http://www.worldcat.org/isbn/0750626348} {\emph
  {\bibinfo {title} {Electrodynamics of Continuous Media}}},\ \bibinfo
  {edition} {2nd}\ ed.\ (\bibinfo  {publisher} {Butterworth-Heinemann},\
  \bibinfo {year} {1984})\BibitemShut {NoStop}%
\bibitem [{\citenamefont {Ronning}\ \emph {et~al.}(2008)\citenamefont
  {Ronning}, \citenamefont {Kurita}, \citenamefont {Bauer}, \citenamefont
  {Scott}, \citenamefont {Park}, \citenamefont {Klimczuk}, \citenamefont
  {Movshovich},\ and\ \citenamefont {Thompson}}]{Ronning2008}%
  \BibitemOpen
  \bibfield  {author} {\bibinfo {author} {\bibfnamefont {F.}~\bibnamefont
  {Ronning}}, \bibinfo {author} {\bibfnamefont {N.}~\bibnamefont {Kurita}},
  \bibinfo {author} {\bibfnamefont {E.~D.}\ \bibnamefont {Bauer}}, \bibinfo
  {author} {\bibfnamefont {B.~L.}\ \bibnamefont {Scott}}, \bibinfo {author}
  {\bibfnamefont {T.}~\bibnamefont {Park}}, \bibinfo {author} {\bibfnamefont
  {T.}~\bibnamefont {Klimczuk}}, \bibinfo {author} {\bibfnamefont
  {R.}~\bibnamefont {Movshovich}}, \ and\ \bibinfo {author} {\bibfnamefont
  {J.~D.}\ \bibnamefont {Thompson}},\ }\href
  {http://stacks.iop.org/0953-8984/20/i=34/a=342203} {\bibfield  {journal}
  {\bibinfo  {journal} {Journal of Physics: Condensed Matter}\ }\textbf
  {\bibinfo {volume} {20}},\ \bibinfo {pages} {342203} (\bibinfo {year}
  {2008})}\BibitemShut {NoStop}%
\bibitem [{\citenamefont {Kogan}\ \emph {et~al.}(2009)\citenamefont {Kogan},
  \citenamefont {Martin},\ and\ \citenamefont {Prozorov}}]{Kogan2009}%
  \BibitemOpen
  \bibfield  {author} {\bibinfo {author} {\bibfnamefont {V.~G.}\ \bibnamefont
  {Kogan}}, \bibinfo {author} {\bibfnamefont {C.}~\bibnamefont {Martin}}, \
  and\ \bibinfo {author} {\bibfnamefont {R.}~\bibnamefont {Prozorov}},\ }\href
  {<Go to ISI>://WOS:000268617100098 http://prb.aps.org/pdf/PRB/v80/i1/e014507}
  {\bibfield  {journal} {\bibinfo  {journal} {Phys. Rev. B}\ }\textbf {\bibinfo
  {volume} {80}},\ \bibinfo {pages} {014507} (\bibinfo {year}
  {2009})}\BibitemShut {NoStop}%
\bibitem [{\citenamefont {Bonalde}\ \emph {et~al.}(2003)\citenamefont
  {Bonalde}, \citenamefont {Yanoff}, \citenamefont {Salamon},\ and\
  \citenamefont {Chia}}]{Bonalde2003}%
  \BibitemOpen
  \bibfield  {author} {\bibinfo {author} {\bibfnamefont {I.}~\bibnamefont
  {Bonalde}}, \bibinfo {author} {\bibfnamefont {B.~D.}\ \bibnamefont {Yanoff}},
  \bibinfo {author} {\bibfnamefont {M.~B.}\ \bibnamefont {Salamon}}, \ and\
  \bibinfo {author} {\bibfnamefont {E.~E.~M.}\ \bibnamefont {Chia}},\ }\href
  {\doibase 10.1103/PhysRevB.67.012506} {\bibfield  {journal} {\bibinfo
  {journal} {Phys. Rev. B}\ }\textbf {\bibinfo {volume} {67}},\ \bibinfo
  {pages} {012506} (\bibinfo {year} {2003})}\BibitemShut {NoStop}%
\bibitem [{\citenamefont {Cooper}(1996)}]{Cooper1996}%
  \BibitemOpen
  \bibfield  {author} {\bibinfo {author} {\bibfnamefont {J.~R.}\ \bibnamefont
  {Cooper}},\ }\href {\doibase 10.1103/PhysRevB.54.R3753} {\bibfield  {journal}
  {\bibinfo  {journal} {Phys. Rev. B}\ }\textbf {\bibinfo {volume} {54}},\
  \bibinfo {pages} {R3753} (\bibinfo {year} {1996})}\BibitemShut {NoStop}%
\bibitem [{\citenamefont {Prozorov}\ \emph {et~al.}(2003)\citenamefont
  {Prozorov}, \citenamefont {Giannetta}, \citenamefont {Kameda}, \citenamefont
  {Tamegai}, \citenamefont {Schlueter},\ and\ \citenamefont
  {Fournier}}]{Prozorov2003}%
  \BibitemOpen
  \bibfield  {author} {\bibinfo {author} {\bibfnamefont {R.}~\bibnamefont
  {Prozorov}}, \bibinfo {author} {\bibfnamefont {R.~W.}\ \bibnamefont
  {Giannetta}}, \bibinfo {author} {\bibfnamefont {N.}~\bibnamefont {Kameda}},
  \bibinfo {author} {\bibfnamefont {T.}~\bibnamefont {Tamegai}}, \bibinfo
  {author} {\bibfnamefont {J.~A.}\ \bibnamefont {Schlueter}}, \ and\ \bibinfo
  {author} {\bibfnamefont {P.}~\bibnamefont {Fournier}},\ }\href {\doibase
  10.1103/PhysRevB.67.184501} {\bibfield  {journal} {\bibinfo  {journal} {Phys.
  Rev. B}\ }\textbf {\bibinfo {volume} {67}},\ \bibinfo {pages} {184501}
  (\bibinfo {year} {2003})}\BibitemShut {NoStop}%
\bibitem [{\citenamefont {Kim}\ \emph {et~al.}(2012{\natexlab{b}})\citenamefont
  {Kim} \emph {et~al.}}]{KimAl}%
  \BibitemOpen
  \bibfield  {author} {\bibinfo {author} {\bibfnamefont {H.}~\bibnamefont
  {Kim}} \emph {et~al.},\ }\href@noop {} {\bibfield  {journal} {\bibinfo
  {journal} {unpublished}\ } (\bibinfo {year}
  {2012}{\natexlab{b}})}\BibitemShut {NoStop}%
\bibitem [{\citenamefont {Zhao}\ \emph {et~al.}(2012)\citenamefont {Zhao},
  \citenamefont {Lausberg}, \citenamefont {Kim}, \citenamefont {Tanatar},
  \citenamefont {Brando}, \citenamefont {Prozorov},\ and\ \citenamefont
  {Morosan}}]{Zhao2012}%
  \BibitemOpen
  \bibfield  {author} {\bibinfo {author} {\bibfnamefont {L.~L.}\ \bibnamefont
  {Zhao}}, \bibinfo {author} {\bibfnamefont {S.}~\bibnamefont {Lausberg}},
  \bibinfo {author} {\bibfnamefont {H.}~\bibnamefont {Kim}}, \bibinfo {author}
  {\bibfnamefont {M.~A.}\ \bibnamefont {Tanatar}}, \bibinfo {author}
  {\bibfnamefont {M.}~\bibnamefont {Brando}}, \bibinfo {author} {\bibfnamefont
  {R.}~\bibnamefont {Prozorov}}, \ and\ \bibinfo {author} {\bibfnamefont
  {E.}~\bibnamefont {Morosan}},\ }\href {\doibase 10.1103/PhysRevB.85.214526}
  {\bibfield  {journal} {\bibinfo  {journal} {Phys. Rev. B}\ }\textbf {\bibinfo
  {volume} {85}},\ \bibinfo {pages} {214526} (\bibinfo {year}
  {2012})}\BibitemShut {NoStop}%
\bibitem [{\citenamefont {Campbell}(1969)}]{Campbell1969}%
  \BibitemOpen
  \bibfield  {author} {\bibinfo {author} {\bibfnamefont {A.~M.}\ \bibnamefont
  {Campbell}},\ }\href {http://stacks.iop.org/0022-3719/2/i=8/a=318} {\bibfield
   {journal} {\bibinfo  {journal} {Journal of Physics C: Solid State Physics}\
  }\textbf {\bibinfo {volume} {2}},\ \bibinfo {pages} {1492} (\bibinfo {year}
  {1969})}\BibitemShut {NoStop}%
\bibitem [{\citenamefont {Campbell}(1971)}]{Campbell1971}%
  \BibitemOpen
  \bibfield  {author} {\bibinfo {author} {\bibfnamefont {A.~M.}\ \bibnamefont
  {Campbell}},\ }\href {http://stacks.iop.org/0022-3719/4/i=18/a=023}
  {\bibfield  {journal} {\bibinfo  {journal} {Journal of Physics C: Solid State
  Physics}\ }\textbf {\bibinfo {volume} {4}},\ \bibinfo {pages} {3186}
  (\bibinfo {year} {1971})}\BibitemShut {NoStop}%
\bibitem [{\citenamefont {Kogan}\ \emph {et~al.}(2006)\citenamefont {Kogan},
  \citenamefont {Prozorov}, \citenamefont {Bud'ko}, \citenamefont {Canfield},
  \citenamefont {Thompson}, \citenamefont {Karpinski}, \citenamefont
  {Zhigadlo},\ and\ \citenamefont {Miranovic}}]{Kogan2006}%
  \BibitemOpen
  \bibfield  {author} {\bibinfo {author} {\bibfnamefont {V.~G.}\ \bibnamefont
  {Kogan}}, \bibinfo {author} {\bibfnamefont {R.}~\bibnamefont {Prozorov}},
  \bibinfo {author} {\bibfnamefont {S.~L.}\ \bibnamefont {Bud'ko}}, \bibinfo
  {author} {\bibfnamefont {P.~C.}\ \bibnamefont {Canfield}}, \bibinfo {author}
  {\bibfnamefont {J.~R.}\ \bibnamefont {Thompson}}, \bibinfo {author}
  {\bibfnamefont {J.}~\bibnamefont {Karpinski}}, \bibinfo {author}
  {\bibfnamefont {N.~D.}\ \bibnamefont {Zhigadlo}}, \ and\ \bibinfo {author}
  {\bibfnamefont {P.}~\bibnamefont {Miranovic}},\ }\href {<Go to
  ISI>://WOS:000242409100102 http://prb.aps.org/pdf/PRB/v74/i18/e184521}
  {\bibfield  {journal} {\bibinfo  {journal} {Phys. Rev. B}\ }\textbf {\bibinfo
  {volume} {74}},\ \bibinfo {pages} {184521} (\bibinfo {year}
  {2006})}\BibitemShut {NoStop}%
\bibitem [{\citenamefont {Gor'kov}(1960)}]{Gorkov60}%
  \BibitemOpen
  \bibfield  {author} {\bibinfo {author} {\bibfnamefont {L.~P.}\ \bibnamefont
  {Gor'kov}},\ }\href@noop {} {\bibfield  {journal} {\bibinfo  {journal}
  {Soviet Phys. JETP}\ }\textbf {\bibinfo {volume} {10}},\ \bibinfo {pages}
  {593} (\bibinfo {year} {1960})}\BibitemShut {NoStop}%
\end{thebibliography}

%

\end{document}